Title: Extending positive CLASS results across multiple instructors and multiple classes of Modeling Instruction

Eric Brewe[1,2], Adrienne Traxler[2], Jorge de la Garza[2,3], Laird H. Kramer[2]

[1] Department of Teaching and Learning, Florida International University, Miami, FL 33199

[2] Department of Physics, Florida International University, Miami, FL 33199

[3] Departamento de Física, Tecnológico de Monterrey, Campus Monterrey

ABSTRACT: We report on a multi-year study of student attitudes measured with the Colorado Learning Attitudes about Science Survey (CLASS) in calculus-based introductory physics taught with the Modeling Instruction curriculum. We find that five of six instructors and eight of nine sections using Modeling Instruction showed significantly improved attitudes from pre- to post-course. Cohen's $d$ effect sizes range from 0.08 - 0.95 for individual instructors. The average effect was $d = 0.45$, with a 95% confidence interval of (0.26 – 0.64). These results build on previously published results showing positive shifts in attitudes from Modeling Instruction classes. We interpret these data in light of other published positive attitudinal shifts and explore mechanistic explanations for similarities and differences with other published positive shifts.

I. INTRODUCTION

A tremendous amount of productive research has addressed improving the conceptual understanding and problem-solving skills of students in physics courses. More recently, a growing number of studies have considered the roles that attitudes, beliefs, and expectations play in learning. Though less immediately familiar to many physics instructors than evaluation of free-body diagrams or correct use of conservation of energy, attitudinal factors have been recognized as a key developing area of research in STEM education [1-3].

In rough analogy to the fruitful measurement of student conceptual gains by diagnostics such as the Force Concept Inventory [4] or the Force and Motion Conceptual Evaluation [5], several instruments have been developed and used to evaluate students' "belief gains" (or losses) [6-8]. We have chosen to use the Colorado Learning Attitudes about Science Survey (CLASS) [7]. No clear consensus has yet emerged as to what types of instruction may reliably improve students' attitudes toward science from pre- to post-course. In the absence of any massive attitudinal data aggregation showing distinct grouping by instructional style [9], researchers are still seeking common factors in the relatively few classrooms with positive gains to report. Further, in light of Pollock & Finkelstein's [10] work on sustaining educational reforms, this article demonstrates an example of a reform that seems to work consistently across multiple implementations and multiple instructors. We contribute to the ongoing literature investigating mechanistic explanations for attitudinal shifts and characteristics of implementations of transformed curricula that promote enhanced student attitudes.

Brewe et al. [11] and de la Garza & Alarcon [12], both using variants of the Modeling Instruction curriculum, published the first favorable attitudinal shifts in university-level calculus-based introductory courses with single classes. Because these prior results came from single courses, the possibility exists that these are the result of instructor effects. To address this concern, we have undertaken a multi-year follow-up study of attitudinal shifts in the Modeling Instruction sections of calculus-based Physics I at Florida International University. The goal of this paper is threefold: to reiterate the Brewe et al. and de la Garza & Alarcon positive attitudinal shifts across multiple years and instructors, to describe features of the Modeling Instruction curriculum that may promote this pattern of favorable shifts, and to add to the ongoing discussion of conditions that may be necessary or sufficient to cause attitudinal gains in introductory physics. We consider several candidates for mechanistic explanations for positive attitudinal shifts: use of the Modeling Instruction curriculum and pedagogy, instructor participation in weekly planning meetings, the epistemological framing of the curriculum, the size of the class, and even whether the CLASS is measuring elements of self efficacy.

Section II reviews previous results in the literature. Section III describes the relevant features of Modeling Instruction, the curriculum common to the data in this paper. Section IV describes our institutional context and methods of data collection. Section V presents results from nine new classes of Physics I. Section VI discusses possible interpretations of those results, and Section VII presents our final conclusions.

## II. PREVIOUS ATTITUDE RESULTS

Physics instructors care about student attitudes and beliefs for a variety of reasons, both personal and pragmatic. At the large scale, the public's perception of science has far-reaching consequences for funding of research in a democracy. Whether or not students eventually major in science, they are exposed to attitudes toward science and learning science from their K-12 and college instructors. In the college classroom, students' attitudes toward science and math have been shown as key predictors of their success in those classes [13, 14]. Narrowing the scope further to physics, students' pretest performance on a survey of attitudes and beliefs has been linked to their decision to take physics courses, as well as their conceptual gains once they do [8, 15]. Longitudinal work using the CLASS shows a strong connection between students' initial attitudes and their persistence in the physics major [16].

One subset of the broad category of "attitudes and beliefs", that of epistemological beliefs, warrants further detail here. The term "beliefs" itself carries some theoretical implications. Researchers investigating student epistemologies differ on whether those epistemologies take the form of broad developmental stages, domain-specific but coherent beliefs, or context-sensitive activations of resources (see Elby [3] for expansion of these distinctions). However, across these theoretical commitments, there is evidence that students' understanding of the nature of knowledge and learning can influence their academic development in a variety of ways. Schommer [17] classifies student beliefs into categories about the certainty, source, and nature of knowledge. In a study of students' comprehension of a mathematical passage, Schommer et al. [18] found that a belief in "simple knowledge" (the idea that knowledge is a set of discrete, unambiguous facts) predicted poor performance on a mastery test, as well as overconfidence in that performance. More generally, we might expect students' epistemological beliefs to affect their choice of study strategies and ability to gauge their own learning. For example, if a student views physics as a series of disconnected definitions and math problems, he or she might be completely satisfied with a learning strategy of memorizing equations without thinking about their interrelationships. Some evidence to support such a causal link exists in case study work by Lising and Elby [19], where they use a resources perspective. Lising and Elby found that students' ability to make effective use of physics knowledge they demonstrably possess is mediated by their deployed epistemological resources about where and how that knowledge is applicable.

Both the "beliefs" and "resources" perspectives on student epistemologies are found in physics education research. They share a broad end goal of promoting a view of physics as a coherent system of knowledge, produced by and accessible through systematic reasoning. A growing body of evidence demonstrates that understanding students' attitudes and epistemologies and how they change as a result of instruction is a necessary consideration for any complete picture of science education. When we wish that students would finish our courses better able to "think like scientists," their attitudes and beliefs about the subject form a real and measurable part of that goal-one which can be supported through appropriate curriculum.

Attitudinal shifts in literature

A number of survey instruments have been developed to assess student attitudes and beliefs in physics courses, and to measure the effect that instruction has on these beliefs. The earliest surveys are the Maryland Physics Expectations Survey [6] and the Views About Science Survey (VASS) [20], with later work producing the Epistemological Beliefs Assessment for Physical Science (EBAPS) [21] and the Colorado Learning Attitudes about Science Survey (CLASS) [7]. A common feature of these instruments is that surveyed students show negative attitudinal shifts from the beginning to end of an introductory course. These negative shifts are common even in transformed classes with demonstrated improvements in conceptual learning [6, 7]. Furthermore, students are able to accurately identify favorable or expert-like responses—both before and after instruction. In short, they know how a physicist would answer the survey questions, but they don't identify with those responses in their personal experience [22].

A small but growing number of courses serve as counterpoints to the negative shifts, demonstrating positive shifts in attitudes and beliefs. These results occur for different student populations and on different instruments: in high school physics on the EBAPS and MPEX [23], for pre-service elementary teachers on the CLASS [24, 25], in algebra-based introductory college physics on the MPEX II [24]. Finally, in introductory calculus-based physics, Modeling Instruction courses have demonstrated positive shifts in student beliefs as measured by the CLASS [11, 12]. It is these results, compiled in Table 1 and Figure 1, that we expand upon in this article. We acknowledge that other instruments could have been used, but we have chosen to use the CLASS due to its greater prevalence in the current literature [7, 11, 12, 24, 25].

**Table 1:** Summary of previously published positive CLASS shifts. Standard Errors of the mean are shown in pre, post and shift data, and the 95% Confidence Interval on the effect is shown in the effect size.

| Group | Overall % Favorable Response | | | Effect size (95% C.I. on *d*) |
|---|---|---|---|---|
| | **Pre** | **Post** | **Shift** | |
| Brewe *et al.* 2009 (MI-B) | 68.6 ± 2.8 | 77.5 ± 2.0 | 9.0 ± 2.7 | 0.71 (0.08, 1.30) |
| de la Garza & Alarcon 2010 (MI-G) | 68.4 ± 2.4 | 71.4 ± 2.1 | 3.1 ± 2.2 | 0.21 (-0.21, 0.63) |
| Otero & Gray 2010 (PET) | 53.8 ± 1.2 | 62.6 ± 1.2 | 8.8 ± 1.1 | 0.59 (0.38, 0.80) |
| Lindsey *et al.* 2012 (PbI) | 52.0 ± 0.7 | 60.6 ± 0.7 | 8.6 ± 0.7 | 0.52 (0.40, 0.63) |

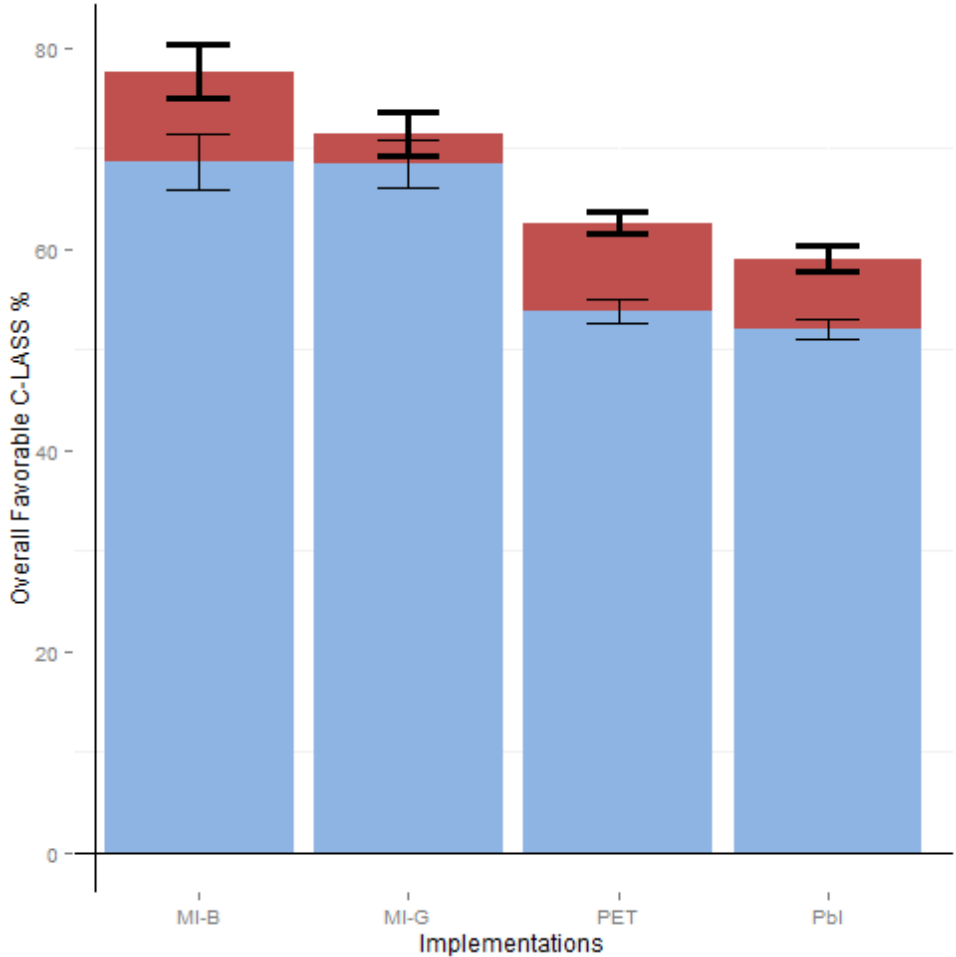

**Figure 1:** Published positive shifts on the CLASS; see also Table 1. Blue and red represent the pre- and post-course percentages, respectively, of overall favorable responses. The error bars show standard error for pre and post.

### III. MODELING INSTRUCTION CURRICULUM AND PEDAGOGY

We contend that the structure of Modeling Instruction (MI) has contributed to the achievement of positive shifts on the CLASS. This structure includes the epistemological foundations which have shaped the development of curricular materials and discourse management practices which ultimately comprise the implementation of MI. Modeling Instruction is a curriculum and pedagogy developed for university physics with an explicit epistemological foundation in the Modeling Theory of Science [27]. According to the Modeling Theory of Science, science is essentially an ongoing process of model development, validation, deployment, and revision. The basic building blocks in science are the models and science is a modeling process. Modeling Instruction contends that science instruction should therefore teach students the basic rules of modeling and should organize the course content around a small set of scientific conceptual models. The scientific conceptual models [28] that structure the course content in MI are shared among members of the learning environment, and the validity, deployment and interpretation of the models are established through classroom activities and discourse.[1] In this approach, models serve as conceptual resources that can be used to develop understanding of a variety of phenomena.

Modeling Instruction's origins in the Modeling Theory of Science provide an epistemological foundation that is evident in the curriculum as student activities are focused on the process of building, validating, and deploying models. This process of modeling replicates the central activity of practicing scientists and therefore promotes students engaging with the practices and norms of physics [29]. The explicit epistemological foundation of physics as model building and use requires that students are active, engaged participants in the learning environment. We contend that the participation of students in model building and use within the Modeling Instruction learning environment contributes to the conceptual and attitudinal gains documented through research [30].

An ongoing, NSF-supported, Modeling Instruction curriculum development and research project has been underway at Florida International University (FIU) since 2007. This effort has built on the success of the High School Modeling Instruction project to develop a set of unique, model-centered curriculum materials appropriate for implementation of Modeling Instruction in university physics. Materials include conceptual reasoning, problem solving, and representational competence building student activities, as well as model building labs. Instructors' materials include guides for implementing the curriculum materials and provide example questions to guide discussion, logistical notes and brief descriptions of the purpose of all student activities. Finally, the curriculum materials include annotated video examples of the materials being used in a Modeling Instruction class. The materials developed by this project were used by all instructors in this study.

Modeling Instruction courses at FIU have primarily operated as a collaborative learning environment, with 30 students in a studio-format class with integrated lab and 'lecture' time. Students in Modeling Instruction classes are active participants, with minimal time spent listening to lectures. Students learn science as scientists, by engaging in inquiry labs, activities targeting model-based reasoning and problem solving; these are the primary mechanisms for building, validating, and using models [27]. In a Modeling Instruction class, the students typically work on an activity in groups of three students and then present their results on a small portable whiteboard to the whole class. In these "board meetings", students discuss their whiteboards and the instructor facilitates and guides the discussion with the goal of helping the class to reach consensus. This modeling cycle builds a view of physics as a coherent system of knowledge where formulas are constructed from conceptual understanding and laboratory evidence, and where models are subject to change as new data become available.

---

[1] Our analysis of models considers scientific conceptual models, which for us are in the shared domain of classroom discourse, rather than mental models, which for us refer to cognitive structures inside minds, and about which we are less certain.

Support for instructors in implementing the Modeling Instruction pedagogy includes an instructors' guide with linked videos and guiding questions for facilitating student discourse. The instructor materials were used as the basis for a weekly Modeling Instruction planning meeting for Instructors 1-4 and 6 represented in this new data. Instructor 5 utilized the materials but was not able to attend the planning meeting. The Modeling Instruction weekly planning meetings lasted one hour and were designed to ensure that the instructors in this data set were using the Modeling Instruction materials and were facilitating student discourse using Modeling Discourse Management practices [31, 32]. During the meetings, the instructors reviewed the instructional plan for the week and reviewed and modified student activities and labs. Reviewing the instructional plan typically involved a discussion of the weekly goals and purposes for specific activities. Due to the attention to goals and purposes, the planning meetings were also a time when the epistemological foundations of the materials were discussed and debated. The planning meetings were typically run by the first author, and were attended by all instructors teaching during any given semester (with the exception of 5).

Several features of the MI learning environment are viable candidates for the influence on attitudes: the active nature of the pedagogy, the explicit epistemological focus on modeling, the small class size, and the effects on students' self efficacy [33]. In addition, the role of the instructor's guide and the weekly planning meetings are worth considering as candidates for the influence on student attitudes. In Section VI we revisit these possibilities in light of accumulated positive CLASS results.

## IV. METHODS

### A. FIU context description

Florida International University is a large urban research-intensive Hispanic-serving institution. As of Fall 2012, enrollment was 50,394 students, 91% of whom are commuters, ensuring that FIU reflects the ethnic diversity of Miami and South Florida. The student body at FIU is 62% Hispanic and 14% Black, making FIU an important source of STEM graduates from underrepresented groups.

The physics department at Florida International University has been experiencing continuous growth in the number of declared and intended physics majors beginning in the early 2000s. The growth represents a nearly 400% increase normalized to the size of the university. Within the physics department several changes are candidates for attribution of this growth, one of which is the implementation of Modeling Instruction in a selection of sections of introductory calculus-based physics. The Modeling Instruction sections of introductory physics are popular, with demand outpacing availability by nearly a factor of four. In order to handle the placement of students in the MI sections the PER group at FIU has implemented a lottery system which both eases administration of registration and provides some randomization of the class makeup.

The student participants in this study are somewhat representative of the FIU student population. Of the 221 participants, 217 reported ethnicity, 76% report Hispanic, 9% Black, 7% White, 6% Asian/Pacific Islander. The sample also includes 120 (47%) male participants and 115 (53%) female participants. This gender distribution is similar to the makeup of the university; however, there is a greater representation of females than in typical physics classes.

B. Pre/post testing in all Modeling Instruction sections of Introductory University Physics

Beginning in Fall 2008 and continuing through Fall 2012, the PER group at FIU has administered the CLASS in all Modeling Instruction sections. The data we report here includes nine sections of the calculus based Mechanics sections (the first semester) of Modeling Instruction. Classes ranged in size from 20 to 30, with an average size of 24.6. We have constrained our analysis to the Mechanics section in order to allow comparisons to other studies with positive attitudinal results. These nine sections include six instructors, two of which have been the instructor for multiple sections. In all cases, the CLASS was administered on the first day of the semester and again on the last day of the semester. The survey was administered as a paper-and-pencil assessment. Student responses were analyzed using the template provided by the University of Colorado PER group[2]. Students who did not participate in both the pre and post test were removed from the data set, as were students who did not respond appropriately to question 31, which indicates if students are reading the questions. Finally, the students in the initial study [11] were not included in this data set to avoid double counting; thus, the data presented here is denoted MI-New. After removing these students, a total of 221 students remained.

Student responses from six different instructors are included in this study. These instructors have a range of experience with University Modeling Instruction. Two instructors (1 & 2) have used University Modeling Instruction curriculum and pedagogy more than five times, two instructors (3 & 4) were implementing for the first time as lead instructor but had each spent a year as an apprentice with experienced Modeling Instruction users, and two instructors (5 & 6) were implementing for the first time without apprentice experience. Instructors 3-6 were all teaching calculus-based introductory physics for the first time, though all had teaching experience in physics labs. Although the results published in Brewe et al. are not included in this current analysis, more recent data from the same instructor are part of this set (as Instructor 1). Further, data from classes taught by Brewe are included in this data set.

In this analysis, we are primarily concerned with the shifts in the overall CLASS profile, and are not including an in-depth analysis of the categories. Also, we are only looking at shifts in favorable responses as has been done in recent analyses [11, 25, 34]. Finally, we calculate effect sizes and confidence intervals on these effects as a way to provide data that are comparable across other studies. The effect size we use is Cohen's *d*, calculated according to Equation 1. The 95% confidence intervals on the effect can be calculated according to Smithson [35].

Equation 1: $$d = \frac{\mu_{post} - \mu_{pre}}{\sigma_{pooled}}$$

Here, μ represents the mean overall percent favorable responses for pre- and post-instruction, and σ is the pooled standard deviation for both sets.

---
[2] Analysis template is available from http://www.colorado.edu/sei/class/

# V. RESULTS AND ANALYSIS

Five of the six instructors who implemented the University Modeling Instruction curriculum and pedagogy achieved significant positive shifts in the overall favorable responses from pre instruction to post instruction, as seen in Figures 2 & 3.

In order to identify the magnitudes of the shifts, we calculated Cohen's *d* and the 95% confidence intervals on the effects for each of the instructors. These are plotted in Figure 2 and range from $d = 0.08$ for instructor 5 to $d = 0.95$ for instructor 4. Based on these effects, students of all instructors – except Instructor 5 – showed positive shifts on the overall CLASS score. Instructor 5 showed no shift as indicated by the effect of $d = 0.08$ and the confidence intervals including $d = 0$. When data from all instructors is aggregated and effect size and confidence intervals on the effect are calculated, we find an effect size of $d = 0.45$, with a 95% confidence interval of (0.26 – 0.64).

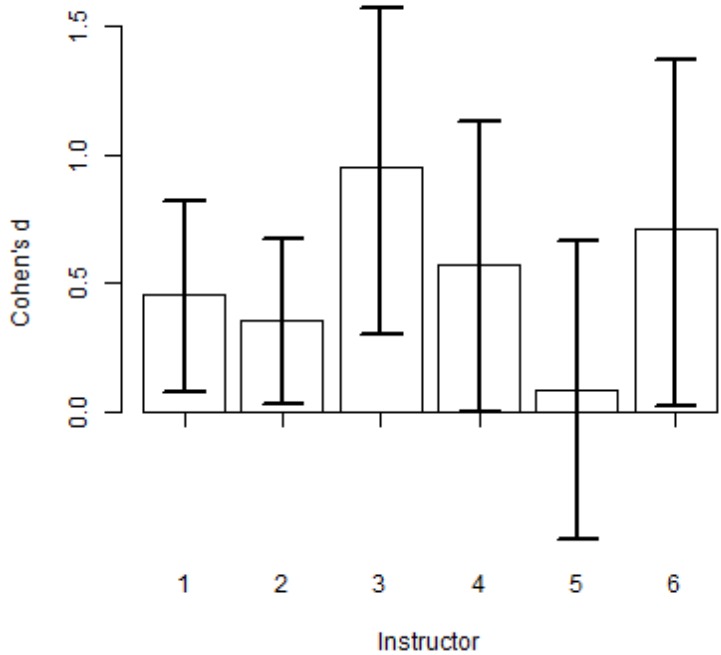

**Figure 2:** Plot of Cohen's *d* by instructor. Error bars reflect the 95% confidence interval on the effect.

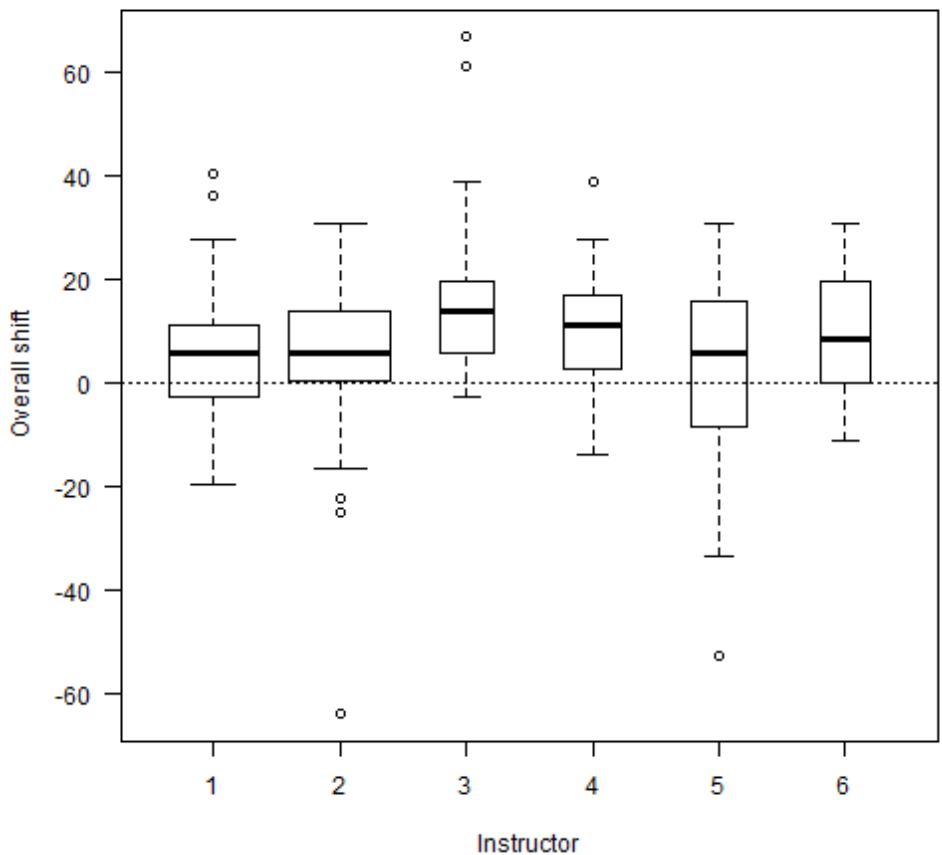

**Figure 3:** Box plot of the shift in overall percent favorable responses for each of the instructors in the new data set. The box width is scaled by the square root of *n* for each instructor. In each box, the thick center line indicates median shift, and the lower and upper bounds represent the first and third quartiles respectively. The whiskers show the extent of the remaining data out to 1.5 times the box size; points outside this range are marked as circles, and represent students with very large shifts.

Although we do not include an analysis of the shifts by category, Appendix A gives the results for all categories as well as the aggregated data by category to allow for meta-analysis of these results. These results indicate significant positive shifts in all categories for aggregated data.

These data do not indicate that all instructors implementing Modeling Instruction would necessarily achieve positive shifts on the CLASS, as Instructor 5 showed no positive shift. We do see consistent positive shifts overall across the majority of instructors, including instructors with varying levels of experience with the University Modeling Instruction curriculum and pedagogy. Further, we see positive shifts across eight of nine sections in this data set. The consistency of these results suggests that the University Modeling Instruction curriculum and pedagogy support the development of more favorable attitudes toward learning physics, independent of instructor. The null result from Instructor 5 warrants further attention.

## VI.     DISCUSSION

The results presented indicate a consistent pattern of positive attitudinal shifts across a variety of instructors. Positive shifts are rare among research reported on calculus-based introductory physics classes. This compilation of positive shifts, especially when contrasted with negative shifts commonly reported, leads us to attribute the shifts to the Modeling Instruction curriculum and pedagogy. The commonalities across these courses provide insight into the factors that could mechanistically explain positive attitudinal shifts. However, the existence of these results alone is not adequate to draw causal conclusions regarding to what specifically the shifts should be attributed. Further, we should attend carefully to Instructor 5 who shared many commonalities with the other instructors, but showed non-significant positive attitudinal shifts.

The most obvious commonality among the instructors is the use of the Modeling Instruction curriculum materials and curriculum guide. Modeling Instruction has bounded this investigation and most clearly distinguishes the participants in this study from others who have not demonstrated positive shifts. Attributing positive shifts to a broad feature such as implementation of a curriculum and pedagogy is unsatisfying, as it does not clearly identify specific features of the implementation that lead to the positive shifts. Yet the curriculum and pedagogy are the features that both unite the instructors in this study and distinguish them from others who have not demonstrated positive attitudinal shifts. In the second half of this section, we provide some suggestions on what aspects of the Modeling Instruction learning environment are particularly relevant to improving student attitudes.

Instructor 5 also implemented the Modeling Instruction curriculum and pedagogy, but as with all implementations, variations naturally occur. One major variation in implementation is that Instructor 5 was unable to participate in the weekly Modeling planning meetings. This could plausibly have had impacts on implementation. Instructor 5 missed out on opportunities to learn from the shared experiences of other instructors and on discussion of the conceptual and epistemological resources that would be valuable in implementation. Either way, the difference is suggestive, and we will in future work pay closer attention to the value of the planning meetings.

A second commonality is that Modeling Instruction at FIU is implemented in classes of 30 students. It is difficult to assume that the small class size, which promotes close faculty/student interaction, does not contribute to improved attitudes. The Modeling classes at FIU are limited to 30 students due to space constraints; however, we anticipate larger sections as classroom space becomes available. Future research will probe whether the positive attitudinal shifts continue in classes two to three times the size of those reported here.

### Considering mechanisms to explain positive shifts

A classroom is a complex learning environment influenced by a multitude of internal and external factors. Claims that a particular measured result is due to any one of these factors, such as curriculum, must be made with great caution. However, by comparing our results with other published studies of students' attitudinal shifts, we can eliminate some factors as likely sources, and strengthen the possibility of attribution to others.

*Unique class/instructor*

The Brewe *et al.* [11] results come from one fall-spring sequence of a course, with one instructor. The MI-New data presented here, spanning five years and five additional instructors (Figure 3), drastically reduce the possibility that positive CLASS shifts arose from a "good semester" or any unique expertise of the professor. Further, these data suggest that the curriculum and the pedagogy that are conveyed by the Modeling Instruction instructors' guide are replicable, even by novice instructors. The role that the weekly planning meetings play in the implementation of the curriculum/pedagogy seems to be important. This preliminary finding fits well with research on dissemination of transformed curricula by Henderson et al. [36] and on the characteristics of high quality professional development [37].

*Class size*

The size element of the instructional environment is less easily dismissible. All of the MI sections in this paper have a maximum enrollment of 30. This small class size is a common feature of several reported positive CLASS results—in Otero, de la Garza et al., and Lindsey et al [12, 25, 34], there is one section of 100 students but all others are 50 students or smaller, with 30 a more typical size. Only Redish and Hammer's positive MPEX-II results come systematically from large courses (100-200 students) [26]. While Redish and Hammer provide counter evidence, it should not be dismissed that the majority of classes showing positive CLASS shifts are small enrollment classes. Some Modeling Instruction curriculum features such as consensus-reaching "board meetings" are currently embedded in the small-class structure; as noted above, it remains to be seen if they will scale successfully to a larger course.

*Epistemological framing of class*

Modeling Instruction is built on an explicit epistemological foundation and the curricular materials and pedagogy are designed to promote the use of productive conceptual and epistemological resources [27]. This epistemological focus is one plausible mechanism for the consistent positive shifts across implementations of Modeling Instruction. In this scenario, students who engage in model building, validation, and revision have authentic scientific experiences. These experiences promote certain attitudes about learning physics: that it is not simply about memorizing formulae, and that models in physics are coherent, constructed by students (and scientists), and subject to change. These attitudes are more aligned with expert attitudes about learning in science, possibly leading to positive shifts on the CLASS.

This scenario is consistent with other positive shifts on the CLASS and other attitude surveys when explicit attention to epistemology was a guiding theme of the course [24, 26]. However, the positive shifts obtained by Lindsey et al. [25] used Physics by Inquiry, which they emphasize does not include any explicit epistemological framing, but instead includes an implicit focus on epistemology. More generally, one aspersion on the epistemological framing argument is that many transformed physics classes have an epistemological component, either explicit or implicit. The prevalence of negative or null shifts in transformed courses, which presumably share epistemological features, is perhaps surprising and a counter argument to the epistemological framing of the class as an explanatory mechanism. We suggest that the theoretical tools available to the physics education research community for characterizing these aspects of curriculum are still developing, and a common language is not yet decided [3]. Documentation of curricular features that address epistemological issues, and how and why they do so, is crucial to furthering understanding of their impact on student attitudes toward learning physics.

*Self-efficacy*

One final, albeit speculative, potential explanatory mechanism is in the CLASS instrument itself. The "learning attitudes" of the CLASS title may be an umbrella term for various, more specific things like expectations, self-efficacy and affect. Self-efficacy is a person's confidence in their ability to accomplish some particular goal (e.g. confidence in their ability to learn physics). This explanation would allow positive shifts to reflect both apprehension prior to the class and then a relief from that apprehension following instruction. Consider the CLASS statement, "A significant problem in learning physics is being able to memorize all the information I need to know." This statement includes an expectation about the nature of learning physics, that "…learning physics is being able to memorize all the information I need to know." It is plausible that this may cause the student to consider her confidence in her ability to memorize all the information, which is a statement of self-efficacy. Thus, an expert-like response on this statement would reflect a combination of expectations about learning and self-efficacy. (This interpretation seems consistent with students' ability to predict expert CLASS responses even when they don't share them [22]—a student might anticipate that a physicist could memorize the requisite information, even if the student could not.) Modeling Instruction has been shown to have either no change or positive shifts on self-efficacy beliefs, whereas standard lecture courses shift self-efficacy beliefs negatively [33, 38]. This pattern of Modeling Instruction demonstrating positive shifts and lecture demonstrating negative shift on self efficacy instruments is reflective of typical CLASS results. This explanation suggests that studying the self-efficacy beliefs embedded in the CLASS statements and how self efficacy is shifted in other classroom contexts (e.g., the Physics by Inquiry setting) is a viable candidate for deeper understanding of CLASS shifts.

## VII. CONCLUSION

We close by returning to the larger picture, considering the relevance of attitudinal results in physics education research. In this subfield, as in education more generally, the past decades have seen growing awareness that no list of facts can encompass mastery of a discipline. But even deeper conceptual understanding, while necessary and desirable for that goal, presents an incomplete picture. Skills such as scientific reasoning, experimental design, and critical evaluation of results have achieved recognition as teachable and measurable aspects of physics. Our results contribute to a growing body of evidence that goes further to address the set of motivations and beliefs that drive and mediate students' learning efforts.

Evidence links student attitudes to their interest and persistence in the discipline (cf. refs 15 and 16). The consistent positive CLASS results at FIU are especially powerful in combination with our concurrent consistent increase in the number of physics majors and the support mechanisms for a student community of physics [30, 40, 41]. We have attempted above to highlight features of Modeling Instruction which may contribute to this pattern of positive shifts, and to rule out some confounding factors. Moving forward, key issues are to explore the question of class size and to continue to continue articulating the epistemological features of classroom and instructional environments.

## VIII. ACKNOWLEDGEMENTS

This work has been supported by the National Science Foundation, (NSF #0802184 , DUE #1140706) and by the Howard Hughes Medical Institute award # 52006924. We also thank Vashti Sawtelle, Renee Michelle Goertzen, Idaykis Rodriguez, Jared Durden and Sean Stewart for their efforts developing the curriculum materials and instructors guides. Finally, we acknowledge the work of Dwain Desbein as a curriculum development partner, and the constructive feedback of two referees.

# APPENDIX A

Although CLASS statistics by category are not essential to the main point of this paper, they are commonly reported in the literature for the CLASS instrument. We include them here to facilitate meta-analysis and comparison with other results.

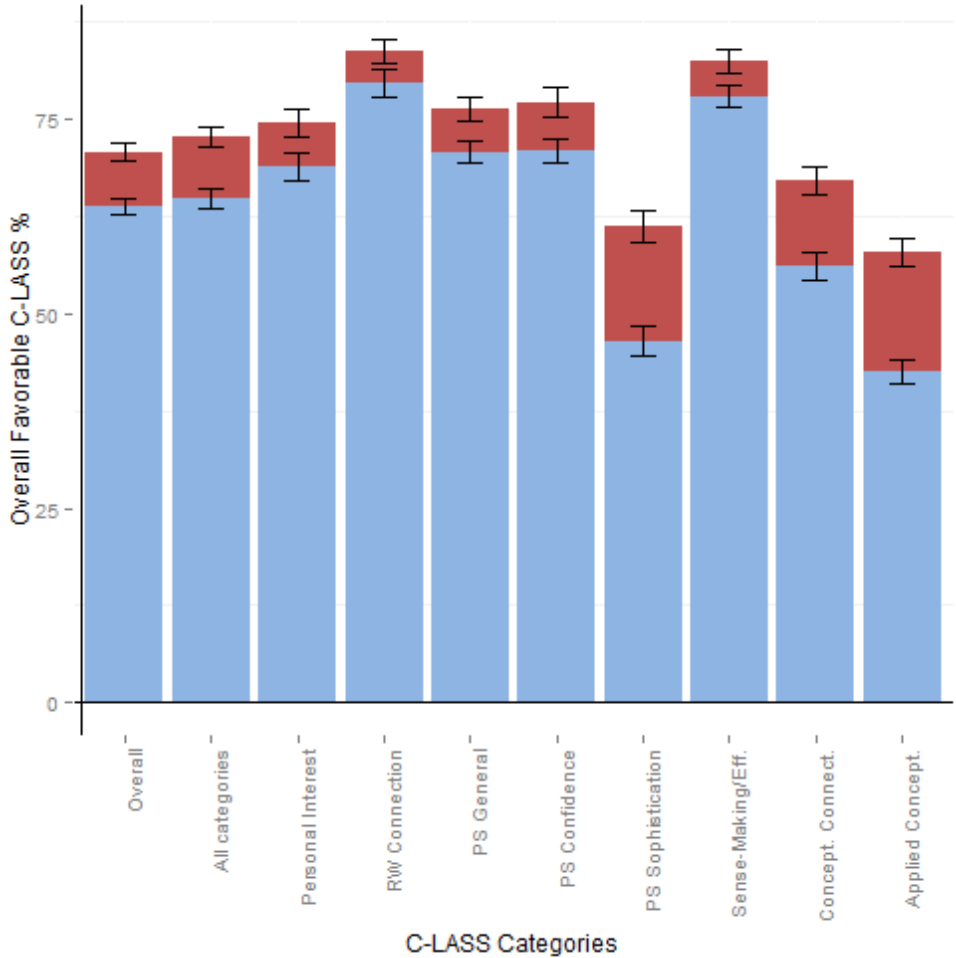

**Figure A1:** MI-New CLASS data (N = 221) for all categories. Blue and red represent the pre- and post-course percentages, respectively, of overall favorable responses. The error bars show standard error for pre and post.

**Table A1:** MI-New CLASS data (N = 221) for all categories ± Standard Error of the mean.

| Category | Pre | Post | Shift |
|---|---|---|---|
| Overall | 63.8 ± 1.0 | 70.8 ± 1.1 | 6.9 ± 1.0 |
| All categories | 64.8 ± 1.2 | 72.8 ± 1.3 | 7.9 ± 1.2 |
| Personal Interest | 68.9 ± 1.8 | 74.5 ± 1.8 | 5.6 ± 1.7 |
| Real-World Connection | 79.7 ± 1.8 | 83.8 ± 1.6 | 4.2 ± 1.9 |
| Problem Solving--General | 70.8 ± 1.4 | 76.4 ± 1.6 | 5.5 ± 1.6 |
| Problem Solving--Confidence | 71.0 ± 1.6 | 77.2 ± 1.9 | 5.9 ± 2.0 |
| Problem Solving--Sophistication | 46.5 ± 1.9 | 61.3 ± 2.0 | 14.8 ± 2.0 |
| Sense-Making/Effort | 78.0 ± 1.3 | 82.4 ± 1.5 | 4.3 ± 1.6 |
| Conceptual Connections | 56.1 ± 1.7 | 67.2 ± 1.8 | 11.1 ± 2.0 |
| Applied Conceptual Understanding | 42.6 ± 1.6 | 57.8 ± 1.8 | 15.1 ± 1.9 |